\documentclass[preprint,5p,times,twocolumn]{elsarticle}

\usepackage[latin1]{inputenc}                    
\usepackage{graphicx}                            
\usepackage{latexsym}                            
\usepackage{amsfonts}                            
\usepackage{amssymb}                             
\usepackage{amsmath}                             
\usepackage[mathscr]{eucal}                      
\usepackage{dcolumn}                             
\usepackage{theorem}                             
\usepackage{footnote}
\usepackage{enumitem}
\usepackage{bm}
\usepackage{color}
\usepackage{natbib}
\usepackage[colorlinks=true,linkcolor=blue]{hyperref}

\def\be{\begin{equation}}
\def\ee{\end{equation}}
\def\bea{\begin{eqnarray}}
\def\eea{\end{eqnarray}}

\journal{Physics Letters B}

\begin{document}

\title{Constraints on the $s-\bar s$ asymmetry of the proton
	in chiral effective theory}
	
\author{X.G.~Wang${}^{a}$,
Chueng-Ryong Ji${}^{b}$,
W.~Melnitchouk${}^{c}$,
Y.~Salamu${}^{d}$,
A.W.~Thomas${}^{a}$,
P. Wang${}^{d,e}$\\[5mm]
\itshape{$^a$ CoEPP and CSSM, University of Adelaide,
		Adelaide SA 5005, Australia}\\[0mm]
\itshape{$^b$ North Carolina State University,
		Raleigh, North Carolina 27695, USA}\\[0mm]
\itshape{$^c$ Jefferson Lab, Newport News,
		Virginia 23606, USA}\\[0mm]
\itshape{$^d$ Institute of High Energy Physics, CAS,
		Beijing 100049, China}\\[0mm]
\itshape{$^e$ Theoretical Physics Center for Science Facilities, CAS,
		Beijing 100049, China}
}

\date{\today}

\begin{abstract}
We compute the $s-\bar s$ asymmetry in the proton in chiral effective
theory, using phenomenological constraints based upon existing data.
Unlike previous meson cloud model calculations, which accounted for
kaon loop contributions with on-shell intermediate states alone, this
work includes off-shell terms and contact interactions, which impact
the shape of the $s-\bar s$ difference.
We identify a valence-like component of $s(x)$ which is balanced by a
$\delta$-function contribution to $\bar s(x)$ at $x=0$, so that the
integrals of $s$ and $\bar s$ over the experimentally accessible region
$x > 0$ are not equal.
Using a regularization procedure that preserves chiral symmetry and
Lorentz invariance, we find that existing data limit the integrated
value of the second moment of the asymmetry to the range
$-0.07 \times 10^{-3}
    \leq \langle x(s-\bar s) \rangle \leq
  1.12 \times 10^{-3}$ at a scale of $Q^2=1$~GeV$^2$.
This is too small to account for the NuTeV anomaly and of the wrong
sign to enhance it.
\end{abstract}

\begin{keyword}
Strange asymmetry, chiral symmetry, kaon loops
\end{keyword}

\maketitle


The nature of the quark--antiquark ($q \bar q$) sea, which complements
the three-valence quark structure of the proton, continues to puzzle
and surprise us, as new generations of experiments provide deeper
insights into its dynamical origins.  From the early simple
expectations of a featureless, virtual sea consisting of $q \bar q$
pairs generated by gluon radiation in perturbative quantum
chromodynamics (QCD), a major paradigm shift occurred with
the observation~\cite{NMC, HERMES, NA51, E866} of a
predicted~\cite{Thomas83} large asymmetry between $\bar d$
and $\bar u$ quarks in the proton.
This challenged our traditional view of the nucleon's peripheral
structure, calling into question long held assumptions about the
role of nonperturbative physics in understanding the phenomenology
of parton distribution functions (PDFs).

With the realization that nonperturbative aspects of QCD were vital
for understanding the 5-quark Fock state components of the nucleon
light-front wave function \cite{Thomas83, Speth98, Vogt01, Garvey01,
Chang11}, an obvious question to ask was whether such effects could
lead to other nontrivial features in the $q\bar q$ sea.
An asymmetry between $s$ and $\bar s$ quarks in the nucleon, as
anticipated by Signal and Thomas~\cite{Signal87}, was a natural
consequence of SU(3) chiral symmetry breaking in QCD, and speculation
later also arose about quark--antiquark asymmetries for charm and
heavier quarks~\cite{MTcharm97, Paiva98, Hobbs14, Lyonnet15}.
Similar considerations led to questioning the traditional expectations
of flavor symmetric polarized sea quarks~\cite{Signal91} and even the
assumption of charge symmetry in the nucleon PDFs \cite{Sather92,
Rodionov94, Londergan10}.

Apart from its intrinsic interest, the possible strange quark
asymmetry, $s-\bar s$, is of great importance in connection with
its contribution to the Paschos-Wolfenstein ratio and the NuTeV
anomaly~\cite{NuTeV}, which suggested a surprisingly large value
for the weak mixing angle, $\sin^2\theta_W$.  A positive value
of the integrated difference, or second moment, of the $s-\bar s$ asymmetry
\begin{equation}
S^-\ \equiv\ \langle x(s-\bar s) \rangle\
= \int_0^1 dx\, x\, (s(x)-\bar s(x)),
\label{e.S-}
\end{equation}
of the order $S^- \sim 2 \times 10^{-3}$, along with other corrections
such as charge symmetry violation, was found to significantly reduce
the excess and bring the NuTeV $\sin^2\theta_W$ measurement closer
to the Standard Model value~\cite{Bentz10}.

Unfortunately, a reliable estimate of the strange asymmetry has been
very difficult to obtain.  An analysis of early $\nu$ and $\bar\nu$
deep-inelastic scattering (DIS) data from BEBC, CDHS and CDHSW
\cite{Zomer00} found a harder $s$ distribution compared with $\bar s$,
albeit with a rather large uncertainty,
	$S^- \approx (2 \pm 3) \times 10^{-3}$.
More recent experimental information has come from dimuon production
in neutrino-nucleus reactions at Fermilab by the CCFR~\cite{CCFR}
and NuTeV~\cite{Zeller02} collaborations, with an NLO analysis
finding
	$S^- = (1.96 \pm 1.43) \times 10^{-3}$
at $Q^2 = 16$~GeV$^2$~\cite{Mason07}.

On the theoretical side, calculations based upon fluctuations into
meson-baryon Fock components \cite{Malheiro97, Malheiro99, Cao03,
Barz98, Alwall04, Ma05, Wakamatsu05, Hobbs15} have led to a fairly
wide range of predictions,
	$S^- \sim (-1~\rm to~+9 ) \times 10^{-3}$,
resulting from the ad hoc assumptions of those models.
%
%
Clearly, if one is to make reliable predictions for $S^-$, a more
systematic approach is needed, one which has a more direct connection
to the underlying QCD theory.

In this Letter we present the first systematic chiral treatment of
the $s-\bar{s}$ asymmetry guided by the need to preserve the model
independent leading nonanalytic (LNA) behavior of the moments of
the strange PDFs.  This work builds upon the unambiguous connection
between the kaon cloud of the nucleon and QCD which followed the
realization~\cite{TMS00} that in chiral expansions of moments of
strange quark PDFs, the coefficients of the LNA terms in the
kaon mass $m_K$ are model independent and can only arise from
pseudoscalar meson loops. 
Starting from the most general effective Lagrangian consistent with
the chiral symmetry of QCD, at a given order in the chiral expansion
a unique set of diagrams can be identified and computed 
systematically~\cite{Arndt01, Chen02}.
The long distance ($m_K \to 0$) effects in such expansions are
thus dictated solely by chiral symmetry and gauge invariance,
while the short distance contributions are treated with a
particular regularization procedure.  
The connection with the chiral theory allows us to identify, for the
first time, a $\delta$-function contribution to the $\bar s$ PDF at
$x=0$, as well as a valence-like component of the $s$-quark PDF.
This result complements earlier discussions of $\delta$-function
contributions in the context of the unpolarized Schwinger term
and proton spin sum rules \cite{Broadhurst73, Bass05}.


\label{sec:convolution}
\begin{figure}[tb]
\includegraphics[width=\columnwidth]{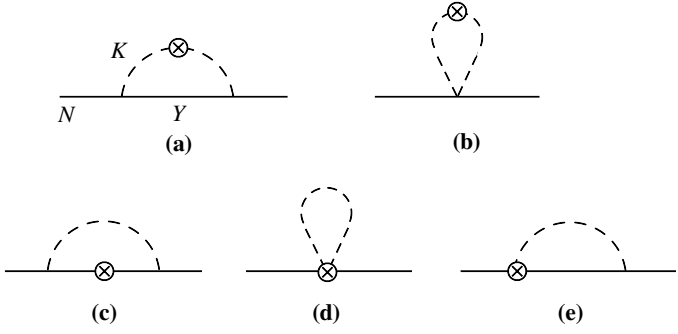}
\caption{Loop contributions to the $\bar s$ PDF from
	(a) kaon rainbow and
	(b) kaon bubble diagrams,
	and to the $s$-quark PDF from
	(c) hyperon rainbow,
	(d) tadpole, and
	(e) Kroll-Ruderman diagrams.
	Nucleons $N$ and hyperons $Y=\Lambda, \Sigma$ are denoted
	by external and internal solid lines, respectively,
	and kaons $K$ by dashed lines, with crosses $\otimes$
	representing insertions of the vector current.
	The Kroll-Ruderman diagram with a current insertion
	on the right-hand vertex is not shown.}
\label{f.loops}
\end{figure}
Expanding the chiral SU(3) Lagrangian to lowest order, the complete
set of diagrams that contribute to $s-\bar s$ is illustrated in
Fig.~\ref{f.loops}.  The direct couplings to the kaon loops in
Fig.~\ref{f.loops}(a) and (b) contribute to the $\bar s$ PDF,
while the $s$-quark PDF contributions arise from the diagrams
involving couplings to hyperons illustrated in
Fig.~\ref{f.loops}(c)--(e).
A general feature of the chiral effective theory constrained analyses
is the presence of contact terms in Figs.~\ref{f.loops}(b) and (d)
that give rise to contributions at zero kaon light-cone momentum
fractions $y = k^+/p^+$, where $k$ is the four-momentum carried by
the kaon and $p$ the four-momentum of the proton.  These are typically
not accounted for in model calculations, which include only the rainbow
diagrams in Figs.~\ref{f.loops}(a) and (c).  The Kroll-Ruderman (KR)
terms represented in Fig.~\ref{f.loops}(e) are needed to preserve
gauge invariance.

The loop contributions to $\bar s$ from the kaon rainbow and kaon
bubble diagrams can be written as a standard convolution of
nucleon $\to$ kaon $+$ hyperon splitting functions,
$f_{KY}^{(\rm rbw)}$ and $f_K^{(\rm bub)}$, with the $\bar s$
PDF in the kaon,
\begin{equation}
\bar{s}(x)\
 =\ \Big( \sum_{KY} f_{KY}^{(\rm rbw)}
        + \sum_K    f_K^{(\rm bub)}
    \Big) \otimes \bar s_K,
\label{e.sbar_conv}
\end{equation}
where the rainbow terms are summed over
  $KY = K^+ \Lambda$, $K^+ \Sigma^0$ and $K^0 \Sigma^+$,
and the kaon bubble terms over
  $K = K^+, K^0$,
and $\otimes$ denotes the convolution \cite{Burkardt13, Salamu15},
$f \otimes q
= \int_0^1 dy \int_0^1 dz$
  \mbox{$f_j(y)\, q(z)\, \delta(x-yz)$}.
The $s$-quark PDF is also a convolution,
\begin{align}
s(x)\ =\
&\sum_{YK}
    \Big( \bar{f}_{YK}^{(\rm rbw)} \otimes s_Y
        + \bar{f}_{YK}^{(\rm KR)}  \otimes s_Y^{(\rm KR)}
    \Big)		 \notag\\
&+\ \sum_K \bar{f}_K^{(\rm tad)} \otimes s_K^{(\rm tad)},
\label{e.s_conv}
\end{align}
where \mbox{$\bar f(y) \equiv f(1-y)$}.
The hyperon rainbow contributions $f_{YK}^{(\rm rbw)}$ are
again summed over all $YK$ combinations, and $f_{YK}^{(\rm KR)}$
are the splitting functions associated with the KR diagrams.
The splitting functions for the tadpole diagram,
Fig.~\ref{f.loops}(d), are equal to the $f_K^{(\rm bub)}$
bubble functions from Fig.~\ref{f.loops}(b).
The strange quark hyperon PDFs $s_Y$, $s_Y^{(\rm KR)}$ and
$s_K^{(\rm tad)}$ for the rainbow, KR and tadpole diagrams,
respectively, can be related to the $u$ and $d$ PDFs in the
proton using SU(3) symmetry.


The splitting function $f_{KY}^{(\rm rbw)}$ in Eq.~(\ref{e.sbar_conv})
for the kaon rainbow diagram can be written as a sum of two terms,
\begin{equation}
f_{KY}^{(\rm rbw)}(y)\
=\ \frac{C_{KY}^2 \overline{M}^2}{(4\pi f_P)^2}
   \left[ f_Y^{\rm (on)}(y) + f_K^{(\delta)}(y)
   \right],
\end{equation}
where $f_Y^{\rm (on)}$ and $f_K^{(\delta)}$ are the on-shell
and $\delta$-function contributions, respectively, $M$ ($M_Y$)
are the nucleon (hyperon) masses, $\overline M = M + M_Y$,
and $f_P$ is the pseudoscalar meson decay constant.
The couplings $C_{KY}$ are given in terms of the SU(3) coefficients
$D$ and $F$.
The on-shell hyperon piece,
\begin{equation}
f_Y^{\rm (on)}(y)\
=\ y \int\!dk_\bot^2\,
   \frac{k_\bot^2 + [M_Y-(1-y) M]^2}{(1-y)^2 D_{KY}^2}
   F^{\rm (on)} \, ,
\label{e.fYon}
\end{equation}
contributes at $y>0$, where
$D_{KY} \equiv -[k_\bot^2 + y M_Y^2 + (1-y) m_K^2 - y(1-y) M^2]/(1-y)$
is the kaon virtuality for an on-shell hyperon intermediate state,
and $F^{\rm (on)}$ is an ultraviolet regulator function.
The function $f_K^{(\delta)}$, on the other hand, arises from
kaons with $y=0$,
\begin{equation}
f_K^{(\delta)}(y)\
=\ \frac{1}{\overline{M}^2}
   \int\!dk_\perp^2\,
   \log\Omega_K\, \delta(y)\,
   F^{(\delta)},
\label{e.fKdel}
\end{equation}      
where $\Omega_K = k_\bot^2 + m_K^2$, and $F^{(\delta)}$ is the
corresponding regulator.
The $K$ bubble diagram in Fig.~\ref{f.loops}(b) originates
with the Weinberg-Tomozawa part of the chiral Lagrangian, and
has a distribution, $f_K^{(\rm bub)}$, similar to the
$\delta$-function part of the rainbow contribution, but with a 
normalization that is independent of the SU(3) couplings,
\begin{equation}
f_{K^+}^{(\rm bub)}\ =\ 2 f_{K^0}^{(\rm bub)}\
=\ -\frac{\overline{M}^2}{(4\pi f_P)^2} f_K^{(\delta)}.
\label{e.fbub}
\end{equation}

For the splitting function associated with the hyperon rainbow
contribution in Eq.~(\ref{e.s_conv}) one finds
\begin{equation}
f_{YK}^{(\rm rbw)}(y)\
=\ \frac{C_{KY}^2 \overline{M}^2}{(4\pi f_P)^2}
   \left[ f_Y^{\rm (on)}(y) + f_Y^{(\rm off)}(y) - f_K^{(\delta)}(y)
   \right],
\end{equation}
where the first (on-shell) and third ($\delta$-function) terms are as
in the kaon rainbow contributions, and the hyperon off-shell term is
\begin{equation}
f_Y^{\rm (off)}(y)\
=\ \frac{2}{\overline{M}} \int\!dk_\bot^2\,
   \frac{\left[M_Y - (1-y) M\right]}{(1-y) D_{KY}}
   F^{\rm (off)} \, ,
\label{e.fYoff}
\end{equation}
with $F^{\rm (off)}$ the corresponding off-shell regulating function.
For the KR contributions in Fig.~\ref{f.loops}(e), necessary for
the preservation of gauge symmetry \cite{JMT13}, one has
\begin{equation}
f_{YK}^{(\rm KR)}(y)\
=\ \frac{C_{KY}^2 \overline{M}^2}{(4\pi f_P)^2}
   \left[ - f_Y^{(\rm off)}(y) + 2 f_K^{(\delta)}(y)
   \right],
\label{e.fKR}
\end{equation}
so that the rainbow and KR contributions satisfy
$f_{YK}^{(\rm rbw)} + f_{YK}^{(\rm KR)} = f_{KY}^{(\rm rbw)}$.
Finally, the tadpole contribution in Fig.~\ref{f.loops}(d)
is related to the bubble term in Eq.~(\ref{e.fbub}),
$f_K^{(\rm tad)} = f_K^{(\rm bub)}$.
These two conditions guarantee that the net strangeness in the
nucleon is zero, \mbox{$\langle s-\bar s \rangle = 0$}.

To regulate the ultraviolet divergences in the splitting functions
one introduces a regularization procedure, such as a cutoff
\cite{Salamu15} or a phenomenological form factor~\cite{Holtmann96}.
Physically, this takes into account the finite size of the baryon
to which the chiral field couples~\cite{Donoghue99, FRR}.
Here we adopt the Pauli-Villars (PV) method, which preserves the
required symmetries and offers many of the advantages of finite
range regularization.
In this approach one subtracts from the point-like amplitudes
expressions in which the propagator mass is replaced by a cutoff
mass $\mu_1$, so that at large momenta the difference between
the amplitudes vanishes \cite{Long16}.
For the $\delta$-function term, because both the $k^-$ and $k_\bot^2$
integrations are divergent, a second subtraction, with regulator 
mass $\mu_2$, is necessary to render the integrals finite.


For the valence PDFs of the mesons we use the recent fit by
Aicher {\it et al.}~\cite{Aicher10}, assuming
\begin{equation} 
  \bar s_{K^+}\ =\ \bar s_{K^0}\ =\ \bar d_{\pi^+}\ .
\end{equation}
The strange quark PDFs in the hyperons are related using SU(3)
symmetry to the $u$ and $d$ PDFs in the proton,
\begin{align}
  s_\Lambda\    &=\ \frac{1}{3}(2 u - d),	\\
  s_{\Sigma^+}\ &=\ s_{\Sigma^0}\ =\ d,
\label{eq:sYsu3}
\end{align}
for which we use parametrization of Martin {\it et al.}~\cite{MRST98}.
For the KR diagrams, the strange PDFs at the $KNY$ vertex
are spin dependent.  They arise because the KR term, which is
required by gauge invariance in the pseudovector chiral theory,
involves pion emission or absorption at the vertex which introduces
a $\gamma^+ \gamma_5$ coupling.
At leading order, SU(3) symmetry requires that these spin dependent
PDFs in the proton are related to the spin-dependent PDFs in the
proton,
\begin{align}
  s^{(\rm KR)}_\Lambda\
&=\ \frac{1}{3F+D}(2 \Delta u - \Delta d),	\\
  s^{(\rm KR)}_{\Sigma^+}\
&=\ s^{(\rm KR)}_{\Sigma^0}\ =\ \frac{1}{F-D} \Delta d.
\label{eq:sYKRsu3}
\end{align}
The fit from Leader {\it et al.} \cite{LSS10} is used for both the
polarized PDFs and the $D$ and $F$ values to ensure each of the
PDFs is normalized to unity.
Given the potentially significant violations of SU(3) symmetry
found in Ref.~\cite{Bass10}, we note that there may be corrections
to the SU(3) PDF relations (\ref{eq:sYsu3}) -- (\ref{eq:sYKRsu3})
at the 10\%--20\% level.
Finally, for the strange PDF at the $ppKK$ vertex of the tadpole
diagram, one has
\begin{align}
s^{(\rm tad)}_{K^+}\ &=\ \frac{1}{2} u,		\\
s^{(\rm tad)}_{K^0}\ &=\ d.
\end{align}
With these relations, the only free parameters in the calculation
are the cutoffs $\mu_1$ and $\mu_2$, which can be constrained
phenomenologically.


The ideal process for constraining $\mu_1$
is inclusive $\Lambda$ hadroproduction, $pp \to \Lambda X$.
At small values of $y$ and $k_\perp$ the $K$ exchange contribution
in Fig.~\ref{f.loops}(a) is expected to dominate, while at higher
momenta heavier meson and baryon intermediate states, as well as
multi meson-exchange processes, will become more important
\cite{Holtmann96}.

\begin{figure}[t]
\includegraphics[width=\columnwidth]{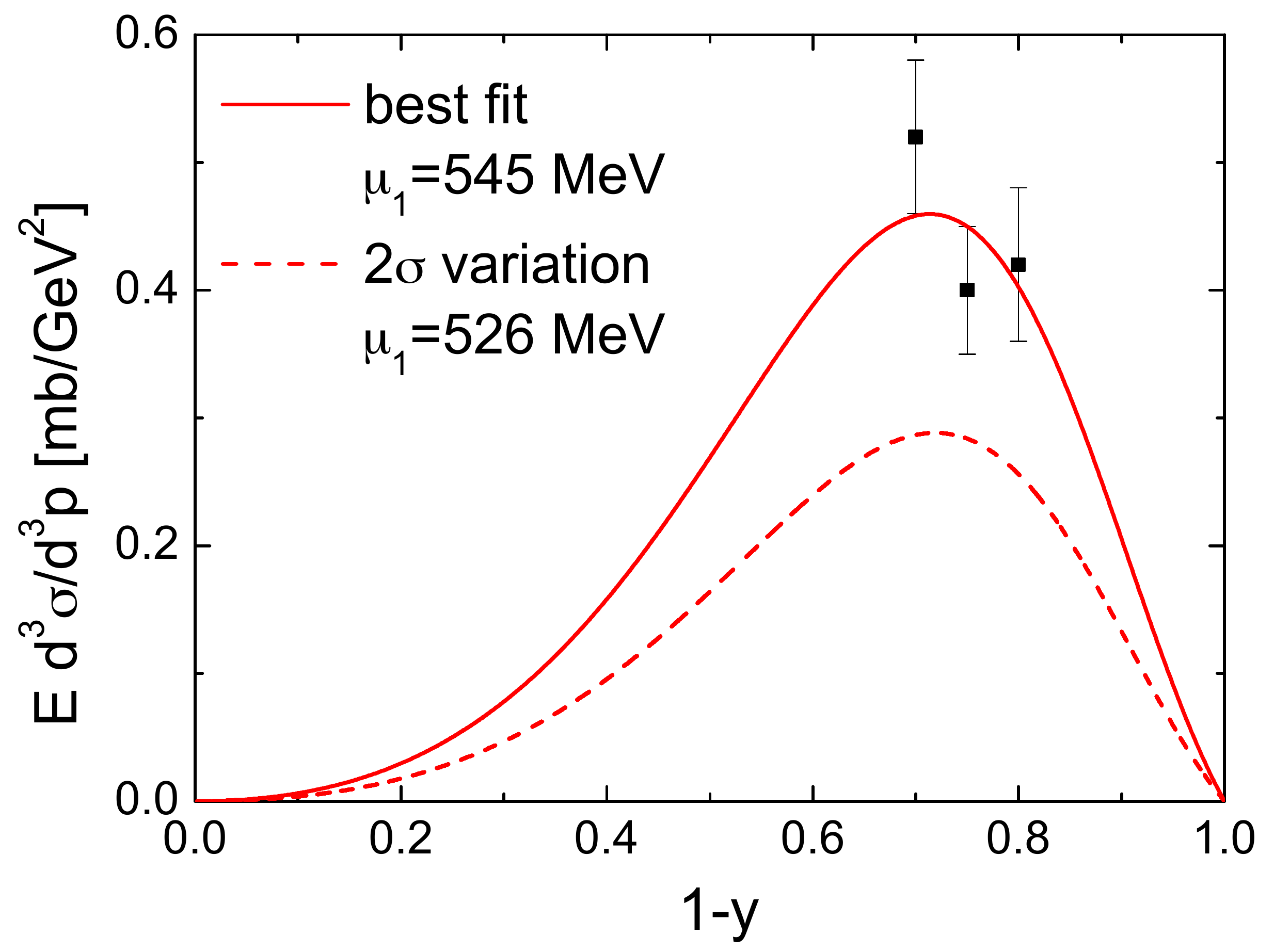}
\caption{Differential cross section for the best fit to the
	$p p \to \Lambda X$ data \cite{Blobel78} in the region
	$y < 0.35$ (solid curve, $\mu_1=545$~MeV),
	as a function of $1-y$ for $k_\perp = 75$~MeV,
	and for a fit $2\sigma$ below the central values
	(dashed curve, $\mu_1=526$~MeV).}
\label{f.fit-Lambda}
\end{figure}

In Fig.~\ref{f.fit-Lambda} we compare the available bubble chamber
data from the CERN proton synchrotron~\cite{Blobel78} for the lowest
available transverse momentum bins.  For the differential cross
section here the current operator corresponds to the total $p K^+$
cross section, for which we take the constant value
$\sigma^{pK^+}_{\rm tot} = (19.9 \pm 0.1)$~mb \cite{Povh92}.
We find the best fit value for the cutoff $\mu_1 = 545$~MeV,
which is taken to yield an upper limit on the kaon contribution.
Contributions from non-kaonic backgrounds may reduce this upper
limit, although at these kinematics the effect should not be large.
As a conservative estimate of the impact of this uncertainty,
we also consider the fit that is two standard deviations lower,
which corresponds to $\mu_1 = 526$~MeV.  These limits yield a
range of momentum fractions carried by $\bar s$ quarks in the
nucleon from
  $\langle \bar s \rangle = 3.4 \times 10^{-3}$ to
			   $5.7 \times 10^{-3}$.

Because the convolution in Eq.~(\ref{e.sbar_conv}) transforms the
$y=0$ contribution in $f_K^{(\delta)}$ to $x=0$, in practice the
$\bar s$ distribution will not provide information on the cutoff
$\mu_2$.  For the $s$-quark PDF, since the convolution in
Eq.~(\ref{e.s_conv}) is expressed in terms of the
splitting functions evaluated at $1-y$, the $f_K^{(\delta)}$
contributions here will be transformed to nonzero values of $x$
and appear valence-like.  Comparison with the $x$ dependence of
the $s$ PDF can then constrain the value of $\mu_2$.

\begin{figure}[t]
\includegraphics[width=\columnwidth]{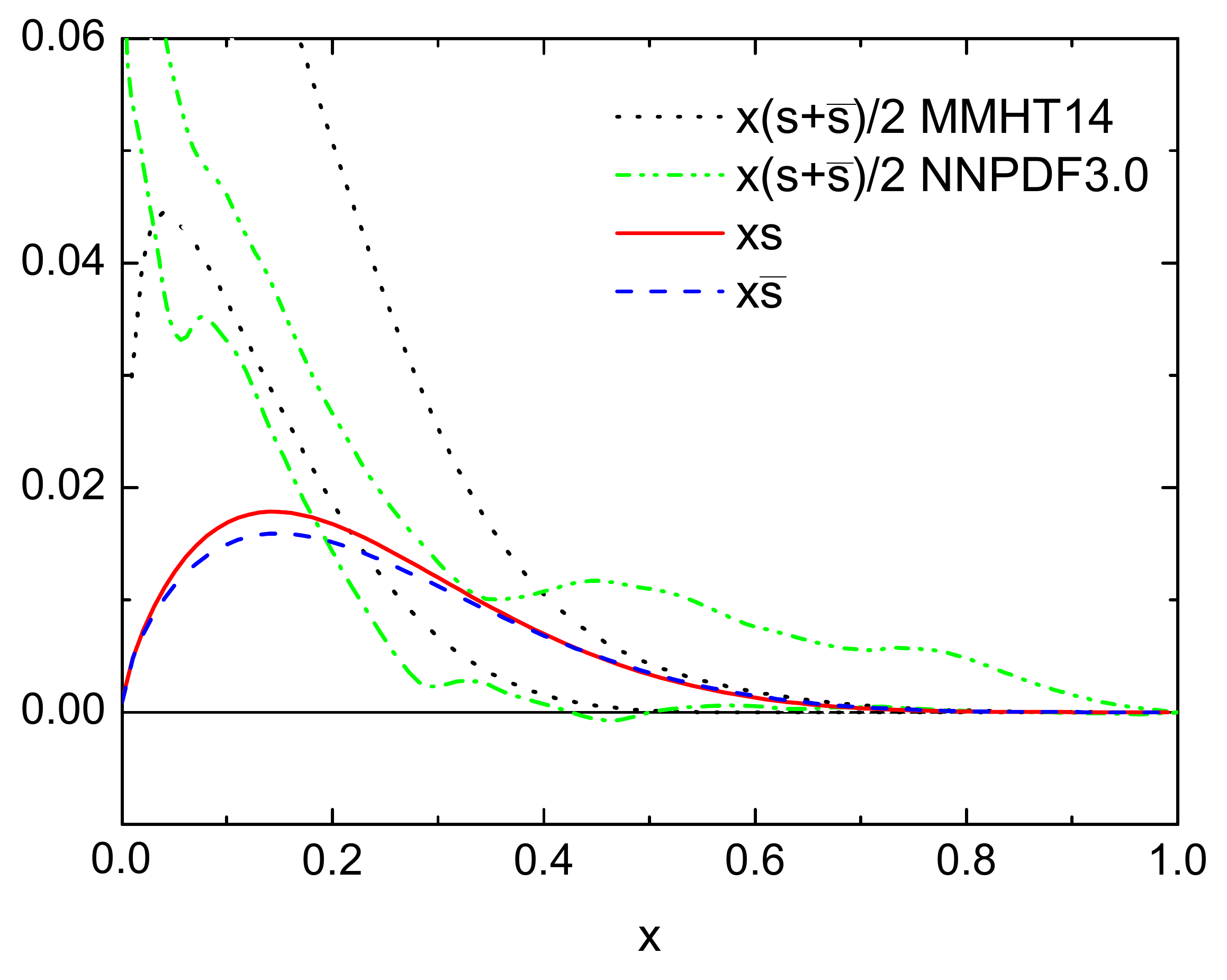}
\caption{Comparison between the strange $xs$ (solid red curve)
	and antistrange $x\bar s$ (dashed blue curve) PDFs
	from kaon loops, for the cutoff parameters
	($\mu_1 = 545$~MeV and $\mu_2 = 600$~MeV) that give
	the maximum total $s+\bar s$, with the upper and lower
	limits of the error bands for $x(s+\bar s)/2$ at
	$Q^2=1$~GeV$^2$ from the MMHT14~\cite{MMHT14} (black dotted)
	and NNPDF3.0~\cite{NNPDF3.0} (green dot-dashed) global fits.}
\label{f.PDFs}
\end{figure}

Our strategy is to fix $\mu_1$ to the maximum value allowed by
the comparison with the $\Lambda$ production data and obtain the
corresponding maximum value for $\mu_2$ such that the calculated
$s+\bar s$ does not exceed the errors on the total phenomenological
PDFs,
  \mbox{$(s+\bar s)_{\rm loops} \leq (s+\bar s)_{\rm tot}$}.
This is illustrated in Fig.~\ref{f.PDFs}, where the individual
$xs$ and $x\bar s$ PDFs from $K$ loops are compared with the recent
average $x(s+\bar s)/2$ parametrization from Refs.~\cite{MMHT14,
NNPDF3.0}.  We see that the calculated curves lie below the
maximum phenomenological values estimated by both the MMHT and
NNPDF collaborations.

For a fixed $\mu_1$, the allowed range for $\mu_2$ with the
PV regularization is
  $m_K \leq \mu_2 \leq \mu_2^{\rm max}$.
At the preferred value found in Fig.~\ref{f.fit-Lambda},
$\mu_1 = 545$~MeV, the upper limit on $\mu_2$ is
$\mu_2^{\rm max} = 600$~MeV.
The corresponding range for the strange asymmetry is
  $-0.07 \times 10^{-3} \leq S^- \leq 0.42 \times 10^{-3}$
at $Q^2=1$~GeV$^2$.
Using the lower value, $\mu_1 = 526$~MeV, reduces the allowed 
momentum that the $s$ quark can carry, and consequently permits
a higher upper limit on $\mu_2$ that still satisfies the
constraint in Fig.~\ref{f.PDFs}.  The limit in this case
becomes $\mu_2^{\rm max} = 894$~MeV, and the range for the
strange asymmetry is
  $-0.01 \times 10^{-3} \leq S^- \leq 1.12 \times 10^{-3}$.
Combining these limits, the strange asymmetry for the maximum
allowed variations on $\mu_1$ and $\mu_2$ consistent with the
available data lies in the range
  $-0.07 \times 10^{-3} \leq S^- \leq 1.12 \times 10^{-3}$.

\begin{figure}[t]
\includegraphics[width=\columnwidth]{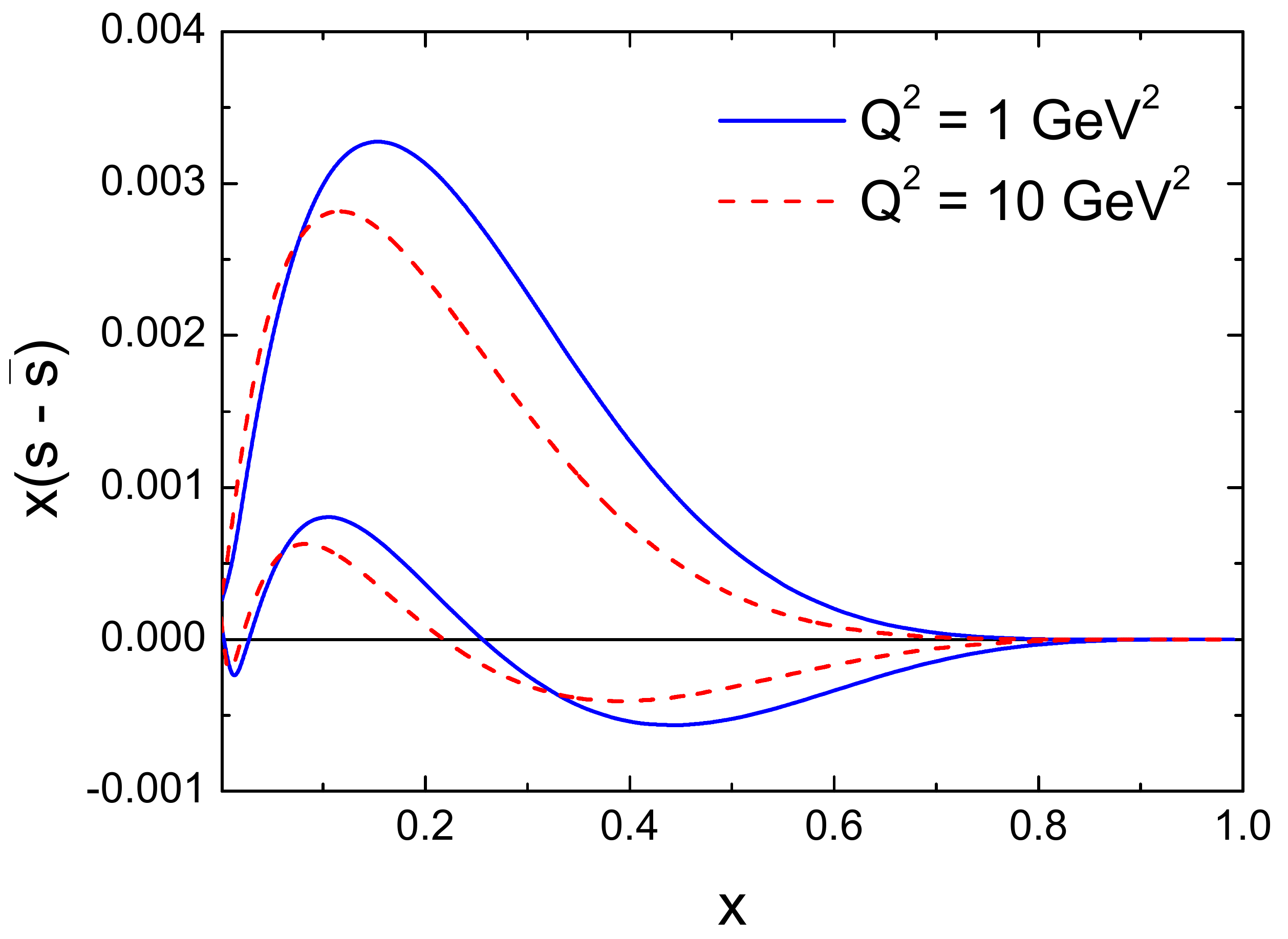}
\caption{Strange quark asymmetry $x(s-\bar s)$ at $Q^2 = 1$~GeV$^2$ 
	(solid blue curves) and evolved to $Q^2 = 10$~GeV$^2$
	(dashed red curves).  The upper (lower) curves correspond
	to the maximum (minimum) value for
	$S^- = 1.12 \times 10^{-3}$
	    ($-0.07 \times 10^{-3}$), for cutoff parameters
	 $\mu_1=526$~MeV, $\mu_2=894$~MeV
	($\mu_1=545$~MeV, $\mu_2=m_K$).}
\label{f.strange-evolution} 
\end{figure}

For these extremal $S^-$ values, the corresponding shape of
$x(s-\bar s)$ is displayed in Fig.~\ref{f.strange-evolution}.
For $\mu_1 = 526$~MeV, the asymmetry remains positive for all $x$,
peaking at $x \approx 0.15$.  Interestingly, for this case there
is no zero crossing at $x > 0$; conservation of strangeness is
ensured by the presence of the nonzero contributions from the
$\delta$-function term $f_K^{(\delta)}$ at $x=0$.
This feature is not present in previous loop calculations
based on kaon loops, which include only rainbow diagrams,
nor in phenomenological PDF fits.
For the parameters that give the minimal $S^-$ value, the $x(s-\bar s)$
distribution also peaks at $x \approx 0.1$, but has a significantly
smaller magnitude.  Furthermore, the distribution becomes negative
for $x \gtrsim 0.2$, which leads to the strong cancellation with the
positive distribution at smaller $x$.

To assess the impact of these asymmetries on the NuTeV anomaly and
the extraction of the weak mixing angle, we fold the calculated
distributions with the acceptance functional for the NuTeV 
data~\cite{Zeller02}.  Varying the $\mu_1$ and $\mu_2$ parameters
over their maximally allowed range, we find a correction,
$\Delta(\sin^2\theta_W)$, to the weak angle from the strange
asymmetry of
     $-7.7 \times 10^{-4} \leq
   \Delta(\sin^2\theta_W) \leq
      -6.7 \times 10^{-7}$
at $Q^2=10$~GeV$^2$.  Remarkably, for all acceptable values of the
cutoff parameters, the correction $\Delta(\sin^2\theta_W)$ remains
negative.  While this has the same sign as that needed to reduce the
NuTeV discrepancy, the small numerical values that we find reduce
the NuTeV anomaly by less than 0.5~$\sigma$.
Had the $S^-$ contribution been large and negative, it would have
enhanced the NuTeV anomaly and further underscored the possibility
of physics beyond the Standard Model.

We have also considered contributions to the asymmetry from kaon
loops accompanied by decuplet hyperons, such as the $\Sigma^*$.
Any contribution to $S^-$ from these is tempered by the need to
reduce the cut-off for the octet component so that the constraint
on $s + \bar s$ is still respected.
As a result, for the range of PV cutoffs considered here 
we find the net effect of the decuplet to be rather small.
Inclusion of higher mass mesons, such as the strange vector $K^*$
mesons~\cite{Cao03, Barz98}, goes beyond the chiral theory framework
and these are more naturally treated as short-distance contributions, 
which should not be added incoherently to other DIS processes.


The virtue of the current study is that we have for the first time
computed the full set of diagrams to lowest order within the effective
chiral theory.
Our analysis has revealed a new contribution to the $\bar s$ PDF
proportional to a $\delta$-function at $x=0$, along with a small
but nonzero valence-like component of the strange PDF.
No phenomenological global PDF fits currently incorporate this
physics, and these may need to be generalized to incorporate more
flexible parametrizations that allow for such behavior.
With the conservative uncertainties chosen for the parameters, we
believe this is the most reliable estimate to date of the chiral
correction to the NuTeV extraction of $\sin^2\theta_W$ from the
strange quark asymmetry.
Ultimately, $s-\bar s$ should be determined empirically and, in
the absence of high precision $\nu$ and $\bar\nu$ data on protons,
the best hope for better constraints may lie with the associated
production of charm with weak bosons at the LHC \cite{Alekhin15}.

\section*{Acknowledgements}

We acknowledge helpful discussions with J.~T.~Londergan at an early
stage of this work.
This work was supported by the DOE Contract No.~DE-AC05-06OR23177,
under which Jefferson Science Associates, LLC operates Jefferson Lab,
DOE Contract No.~DE-FG02-03ER41260, the Australian Research Council
through the ARC Centre of Excellence for Particle Physics at the
Terascale (CE110001104), an ARC Australian Laureate Fellowship
FL0992247 and DP151103101, and by NSFC under Grant No.~11475186,
CRC 110 by DFG and NSFC.

%
%



\begin{thebibliography}{99}

\bibitem{NMC}
M.~Arneodo {\it et al.},
Phys. Rev. D {\bf 50}, 1 (1994).

\bibitem{HERMES}  
K.~Ackerstaff {\it et al.},
Phys. Rev. Lett. {\bf 81}, 5519 (1998).

\bibitem{NA51}
A.~Baldit {\it et al.},
Phys. Lett. B {\bf 332}, 244 (1994).

\bibitem{E866}
R.~S.~Towell {\it et al.},
Phys. Rev. D {\bf 64}, 052002 (2001).

\bibitem{Thomas83}
A.~W.~Thomas,
Phys. Lett. B {\bf 126}, 97 (1983).

\bibitem{Speth98}
J.~Speth and A.~W.~Thomas,
Adv. Nucl. Phys. {\bf 24}, 83 (1998).

\bibitem{Vogt01}
R.~Vogt,
Prog. Part. Nucl. Phys. {\bf 45}, S105 (2000).

\bibitem{Garvey01}
G.~T.~Garvey and J.~C.~Peng,
Prog. Part. Nucl. Phys. {\bf 47}, 203 (2001).

\bibitem{Chang11}
W.-C.~Chang and J.-C.~Peng,       
Phys. Rev. Lett. {\bf 106}, 252002 (2011);
%
J.-C.~Peng and J.-W.~Qiu,
Prog. Part. Nucl. Phys. {\bf 76}, 43 (2014).

\bibitem{Signal87}
A.~I.~Signal and A.~W.~Thomas,
Phys. Lett. B {\bf 191}, 205 (1987).

\bibitem{MTcharm97}
W.~Melnitchouk and A.~W.~Thomas,
Phys. Lett. B {\bf 414}, 134 (1997).

\bibitem{Paiva98}
S.~Paiva, M.~Nielsen, F.~S.~Navarra, F.~O.~Duraes and L.~L.~Barz,
Mod. Phys. Lett. A {\bf 13}, 2715 (1998).

\bibitem{Hobbs14}
T.~J.~Hobbs, J.~T.~Londergan and W.~Melnitchouk,
Phys. Rev. D {\bf 89}, 074008 (2014).

\bibitem{Lyonnet15}
F.~Lyonnet {\it et al.},
JHEP {\bf 1507}, 141 (2015).

\bibitem{Signal91}
A.~I.~Signal, A.~W.~Schreiber and A.~W.~Thomas,
Mod. Phys. Lett. A {\bf 6}, 271 (1991).

\bibitem{Sather92}
E.~Sather,
Phys. Lett. B {\bf 274}, 433 (1992).

\bibitem{Rodionov94}
E.~N.~Rodionov, A.~W.~Thomas and J.~T.~Londergan,
Mod. Phys. Lett. A {\bf 9}, 1799 (1994).

\bibitem{Londergan10}
J.~T.~Londergan, J.~C.~Peng and A.~W.~Thomas,
Rev. Mod. Phys. {\bf 82}, 2009 (2010).

\bibitem{NuTeV}
G.~P.~Zeller {\it et al.},
Phys. Rev. Lett. {\bf 88}, 091802 (2002).

\bibitem{Bentz10}
W.~Bentz, I.~C.~Cloet, J.~T.~Londergan and A.~W.~Thomas,
Phys. Lett. B {\bf 693}, 462 (2010).

\bibitem{Zomer00}
V.~Barone, C.~Pascaud and F.~Zomer,
Eur. Phys. J. C {\bf 12}, 243 (2000).

\bibitem{CCFR}
A.~O.~Bazarko {\it et al.},
Z. Phys. C {\bf 65}, 189 (1995).

\bibitem{Zeller02}
G.~P.~Zeller {\it et al.},
Phys. Rev. D {\bf 65}, 111103(R) (2002); 119902(E) (2003).

\bibitem{Mason07}
D.~Mason {\it et al.},
Phys. Rev. Lett. {\bf 99}, 192001 (2007).

\bibitem{Malheiro97}
W.~Melnitchouk and M.~Malheiro,
Phys. Rev. C {\bf 55}, 431 (1997).

\bibitem{Malheiro99}
W.~Melnitchouk and M.~Malheiro,
Phys. Lett. B {\bf 451}, 224 (1999).

\bibitem{Cao03}
F.~G.~Cao and A.~I.~Signal,
Phys. Lett. B {\bf 559}, 229 (2003).

\bibitem{Barz98}
L.~L.~Barz, H.~Forkel, H.~W.~Hammer, F.~S.~Navarra, M.~Nielsen
and M.~J.~Ramsey-Musolf,
Nucl. Phys. {\bf A640}, 259 (1998).

\bibitem{Alwall04}
J.~Alwall and G.~Ingelman,
Phys. Rev. D {\bf 70}, 111505 (2004).

\bibitem{Ma05}
Y.~Ding, R.~G.~Xu and B.~Q.~Ma,
Phys. Lett. B {\bf 607}, 101 (2005).

\bibitem{Wakamatsu05}
M.~Wakamatsu,
Phys. Rev. D {\bf 71}, 057504 (2005).

\bibitem{Hobbs15}
T.~J.~Hobbs, M.~Alberg and G.~A.~Miller,
Phys. Rev. C {\bf 91}, 035205 (2015).

\bibitem{TMS00}
A.~W.~Thomas, W.~Melnitchouk and F.~M.~Steffens,
Phys. Rev. Lett. {\bf 85}, 2892 (2000).

\bibitem{Arndt01}
D.~Arndt and M.~J.~Savage,
Nucl. Phys. {\bf A697}, 429 (2002).

\bibitem{Chen02}
J.-W.~Chen and X.~Ji,
Phys. Rev. Lett. {\bf 87}, 152002 (2001);
{\bf 88}, 249901(E) (2002).

\bibitem{Broadhurst73}
D.~J.~Broadhurst, J.~F.~Gunion and R.~L.~Jaffe,
Annals Phys. {\bf 81}, 88 (1973).

\bibitem{Bass05}
S.~D.~Bass,
Rev. Mod. Phys. {\bf 77}, 1257 (2005).

\bibitem{Burkardt13}
M.~Burkardt, K.~S.~Hendricks, C.-R.~Ji, W.~Melnitchouk and A.~W.~Thomas,
Phys. Rev. D {\bf 87}, 056009 (2013).

\bibitem{Salamu15}
Y.~Salamu, C.-R.~Ji, W.~Melnitchouk and P.~Wang,
Phys. Rev. Lett. {\bf 114}, 122001 (2015).

\bibitem{JMT13}
C.-R.~Ji, W.~Melnitchouk and A.~W.~Thomas,
Phys. Rev. D {\bf 88}, 076005 (2013).

\bibitem{Holtmann96}
H.~Holtmann, A.~Szczurek and J.~Speth,
Nucl. Phys. {\bf A569}, 631 (1996).

\bibitem{Donoghue99}
J.~F.~Donoghue, B.~R.~Holstein and B.~Borasoy,
Phys. Rev. D {\bf 59}, 036002 (1999).

\bibitem{FRR}
R.~D.~Young, D.~B.~Leinweber and A.~W.~Thomas,
Nucl. Phys. Proc. Suppl. {\bf 141}, 233 (2005).

\bibitem{Long16}
X.G.~Wang, C.-R.~Ji, W.~Melnitchouk, Y.~Salamu, A.~W.~Thomas and P.~Wang,
arXiv:1610.03333 [hep-ph].

\bibitem{Aicher10}
M.~Aicher, A.~Sch\"afer and W.~Vogelsang,
Phys. Rev. Lett. {\bf 105}, 252003 (2010).

\bibitem{MRST98}
A.~D.~Martin, R.~G.~Roberts, W.~J.~Stirling and R.~S.~Thorne,
Eur. Phys. J. C {\bf 4}, 463 (1998).

\bibitem{LSS10}
E.~Leader, A.~V.~Sidorov and D.~B.~Stamenov,
Phys. Rev. D {\bf 82}, 114018 (2010).

\bibitem{Bass10}
S.~D.~Bass and A.~W.~Thomas,
Phys. Lett. B {\bf 684}, 216 (2010).

\bibitem{Blobel78}
V.~Blobel {\it et al.},
Nucl. Phys. {\bf B135}, 379 (1978).

\bibitem{Povh92}
B.~Povh and J.~H\"{u}fner,
Phys. Rev. D {\bf 46}, 990 (1992).

\bibitem{MMHT14}       
L.~A.~Harland-Lang, A.~D.~Martin, P.~Motylinski and R.~S.~Thorne,
Eur. Phys. J. C {\bf 75}, 204 (2015).

\bibitem{NNPDF3.0}
R.~D.~Ball {\it et al.},
JHEP {\bf 04} (2015) 040.

\bibitem{Alekhin15}
S.~Alekhin, J.~Bl\"umlein, L.~Caminadac, K.~Lipka, K.~Lohwasser,
S.-O.~Moch, R.~Petti and R.~Placakyte,
Phys. Rev. D {\bf 91}, 094002 (2015).

\end{thebibliography}
\end{document}